# Relativistic quantum teleportation protected by the anti-Unruh effect

**Reza Hamzehofi[1], Davood Afshar[1,2,a] , Mehrzad Ashrafpour[1,2]**

[1]Department of Physics, Faculty of Science, Shahid Chamran University of Ahvaz, Ahvaz, Iran
[2]Center for Research on Laser and Plasma, Shahid Chamran University of Ahvaz, Ahvaz, Iran

**Abstract** The interaction of an accelerated observer with a quantum field can modify the entanglement and quantum information shared between observers, thereby affecting quantum communication in non-inertial frames. We formulate the standard teleportation protocol in a relativistic setting where Alice remains inertial while Rob undergoes uniform acceleration and is modeled as an Unruh–DeWitt detector locally coupled to a massive scalar field. The detector–field interaction initially leads to a loss of entanglement and quantum information shared between Alice and Rob. Remarkably, under acceleration, the anti-Unruh regime can reverse this degradation, leading to an increase in both the shared entanglement and the accessible quantum information. We further find that, in the anti-Unruh regime, the teleportation fidelity is independent of the local free-evolution time of Alice's and Rob's states. At high accelerations, the recovered entanglement and quantum information enhance the teleportation fidelity, which approaches unity under appropriate conditions. These results demonstrate that the anti-Unruh effect can protect and recover quantum entanglement and information in a non-inertial frame by reducing the detector's effective temperature and the associated decoherence.

## 1 Introduction

Relativistic effects have become an important frontier in quantum information science, particularly in situations where quantum systems undergo noninertial motion. The interplay between relativity and quantum correlations has revealed that acceleration can significantly modify entanglement, coherence, and information flow, thereby challenging the assumptions underlying quantum communication protocols originally formulated in inertial frames [1–4]. Understanding how relativistic motion influences quantum information processing is therefore essential, both from a fundamental perspective and for potential future implementations involving accelerated or space-based quantum technologies.

Among relativistic phenomena, the Unruh effect has been extensively studied in the context of quantum information. It predicts that a uniformly accelerated observer perceives the Minkowski vacuum as a thermal state, leading to degradation of quantum correlations shared with inertial observers [5–9]. This thermalization has been shown to adversely affect a variety of quantum information tasks, including entanglement distribution and quantum teleportation, typically resulting in reduced fidelity and loss of information [10–13]. From this viewpoint, acceleration is often regarded as an unavoidable source of noise that limits the performance of relativistic quantum communication protocols.

More recently, however, it has been demonstrated that the response of accelerated quantum systems can exhibit counterintuitive behavior under certain conditions, a phenomenon commonly referred to as the anti-Unruh effect [14–17]. In this regime, increasing acceleration may suppress

**a** e-mail: da_afshar@yahoo.com (corresponding author)

excitation probabilities or reduce effective thermal noise, suggesting that relativistic motion does not necessarily act as a purely detrimental resource. This observation has motivated renewed interest in exploring whether relativistic effects can be harnessed constructively in quantum information processing, rather than being viewed solely as a source of decoherence.

These developments naturally raise a fundamental question: can relativistic effects, and in particular the anti-Unruh phenomenon, be exploited to enhance the performance of quantum communication protocols? Addressing this question requires a careful investigation of how acceleration modifies both the underlying quantum correlations and the operational figures of merit relevant to information transfer. In this work, we focus on the standard quantum teleportation protocol originally introduced in Ref. [18], and analyze how noninertial motion of one of the parties influences the teleportation fidelity and the transfer of quantum information.

The paper is organized as follows. In Sec. 2, we introduce the physical model and the theoretical framework used to describe the accelerated system and its interaction with the field. Sec. 3 presents a detailed analysis of the teleportation fidelity and its dependence on acceleration and system parameters, with particular emphasis on regimes where anti-Unruh behavior leads to nontrivial and potentially advantageous effects. Finally, Sec. 4 summarizes our main results and discusses their implications for relativistic quantum information processing.

## 2 The Unruh–DeWitt detector model

The Unruh–DeWitt (UDW) detector is a two-level quantum system that couples locally to a quantum field and provides an operational framework for probing field properties along a given trajectory [14]. The detector possesses a ground state $|g\rangle$ and an excited state $|e\rangle$, separated by an energy gap $\Omega > 0$. By analyzing transitions between these states induced by the coupling to the field, one can infer physical properties such as vacuum fluctuations, particle content, and thermal effects associated with acceleration, including the Unruh effect.

### 2.1 Interaction Hamiltonian

The interaction Hamiltonian of the detector with a real scalar field $\phi$, expressed in the detector's proper time $\tau$, is [19]

$$H_I(\tau) = \lambda\chi(\tau)\left(\hat{\sigma}^+ e^{i\Omega\tau} + \hat{\sigma}^- e^{-i\Omega\tau}\right)\phi\left(x(\tau)\right), \tag{1}$$

where $\lambda$ is the coupling constant, $\sigma^{\pm}$ are the ladder operators of the detector, $\Omega$ is the detector energy gap, and $\phi\left(x(\tau)\right)$ is the scalar field evaluated along the detector's worldline $x(\tau)$. A smooth Gaussian switching function

$$\chi(\tau) = e^{-\tau^2/2\sigma^2}, \tag{2}$$

is introduced to localize the detector–field interaction in proper time and to avoid unphysical excitations associated with abrupt switching; the parameter $\sigma$ sets the interaction timescale.

## 2.2 Accelerated trajectory

Before introducing the perturbative dynamics, we specify the kinematic and field-theoretic background. For a uniformly accelerated detector with proper acceleration $a$, the worldline in Minkowski spacetime follows a hyperbolic path parametrized by

$$t(\tau) = \mathrm{a}^{-1}\sinh(a\tau), \qquad (3)$$
$$x(\tau) = \mathrm{a}^{-1}\cosh(a\tau).$$

For a massive real scalar field of mass $m$ in (1+1) dimensions, the dispersion relation gives the mode frequency

$$\omega_k = \sqrt{(k^2 + m^2)}. \qquad (4)$$

The mass $m > 0$ is essential: it is precisely the non-zero field mass, combined with the finite switching timescale $\sigma$, that gives rise to the anti-Unruh effect [20, 21]. For a massless field, the standard Unruh result is recovered and the excitation probability increases monotonically with $a$. Throughout this work we use natural units $\hbar = c = 1$.

## 2.3 Perturbative Evolution

In the weak-coupling regime $(\lambda \ll 1)$, the unitary evolution operator of the combined detector–field system is expanded perturbatively as

$$\hat{U} = \hat{\mathbf{1}} - i\int_{-\infty}^{+\infty} d\tau \hat{H}_I(\tau) + O(\lambda^2). \qquad (5)$$

This first-order approximation captures the leading-order effects of acceleration on detector excitation and de-excitation. At first order in $\lambda$, the unitary evolution induces transformations of the detector–field states of the form [14]

$$\hat{U}|g,0\rangle = |g,0\rangle - i\eta_0|e,1_k\rangle + \cdots, \qquad (6)$$
$$\hat{U}|e,0\rangle = |e,0\rangle + i\eta_1\,|g,1_k\rangle + \cdots,$$

where $|1_k\rangle$ denotes the single-particle Minkowski state in mode $k$ produced by the interaction, $\eta_0$ is the excitation amplitude, the probability amplitude for the detector to transition from $|g\rangle$ to $|e\rangle$ by absorbing a field quantum of mode $k$, and $\eta_1$ is the de-excitation amplitude for the reverse process. These amplitudes are determined by the integral

$$\eta_{0,1} = \lambda\int dk \int_{-\infty}^{+\infty} d\tau\ \chi(\tau) e^{\pm i\Omega\tau} u_k[x(\tau)], \qquad (7)$$

where the upper (lower) sign corresponds to $\eta_0\,(\eta_1)$ and $u_k(x)$ is the positive-frequency Minkowski mode function of the massive scalar field. In (1+1)-dimensional Minkowski spacetime, these are complex plane waves of the standard form

$$u_k(x)=\frac{1}{\sqrt{4\pi\omega_k}}exp\left[i\left(kx-\omega_k t\right)\right]. \tag{8}$$

The dependence on the proper acceleration $a$ enters through the worldline $x(\tau)$ [Eq. (3)], the dependence on the switching width $\sigma$ enters through $\chi(\tau)$ [Eq. (2)], and the dependence on the field mass $m$ enters through the mode frequency $\omega_k$ contained in $u_k(x)$ [Eq. (4)]. Both $\eta_0$ and $\eta_1$ are evaluated numerically along the accelerated trajectory for given values of $(\Omega, a, \sigma, m)$.

2.4 Alice–Rob entangled state

We assume that Alice and Rob, with Rob equipped with a UDW detector, initially share the maximally entangled state

$$|\phi\rangle_{A,R}=\frac{1}{\sqrt{2}}\left(|g,g\rangle+|e,e\rangle\right)_{AR}, \tag{9}$$

while the scalar field is prepared in the Minkowski vacuum $|0\rangle_{field}$ . The total initial state is therefore given by the direct product $|\psi_0\rangle=|\phi\rangle_{A,R}\otimes|0\rangle_{field}$. Rob then undergoes uniform acceleration; as a result, the accelerated detector acquires a nonzero probability of excitation $|g\rangle\to|e\rangle$ even though the field is in the vacuum state from the perspective of inertial observers—a manifestation of the Unruh effect. Applying the transformations (6) to the state (9), and normalizing each branch of the superposition individually so that unitarity is preserved exactly within the perturbative truncation, the state evolves as

$$\begin{aligned}|\psi\rangle=\frac{1}{\sqrt{2}}\Bigg\{&\frac{1}{\sqrt{1+|\eta_0|^2}}\left(|g,g\rangle_{AR}|0\rangle_{field}-i\eta_0|g,e\rangle_{AR}|1_k\rangle_{field}\right)+\frac{1}{\sqrt{1+|\eta_1|^2}}\\ &\times\left(|e,e\rangle_{AR}|0\rangle_{field}+i\eta_1|e,g\rangle_{AR}|1_k\rangle_{field}\right)\Bigg\}.\end{aligned} \tag{10}$$

The two terms in braces correspond to the two original Bell branches, $|g,g\rangle$ and $|e,e\rangle$; each branch is normalized independently by the factors $\left(1+|\eta_0|^2\right)^{-1/2}$ and $\left(1+|\eta_1|^2\right)^{-1/2}$, respectively, since the detector–field interaction redistributes probability within each branch (from $|0\rangle_{field}$ into $|1_k\rangle_{field}$) but does not mix the two branches at this order.

2.5 Reduced density matrix

To obtain the density matrix of the system shared between Alice and Rob, we perform a partial trace over the field modes:

$$\hat{\rho}_{A,R} = Tr_{field}\left[|\psi\rangle\langle\psi|\right]. \tag{11}$$

Since $|0\rangle_{field}$ and $|1_k\rangle_{field}$ are orthogonal, the trace eliminates all cross terms between the two field states, while the cross terms between the $|g,g\rangle$ and $|e,e\rangle$ branches (both proportional to $|0\rangle_{field}$) survive with a combined normalization factor. In the computational basis $\{|g,g\rangle,|g,e\rangle,|e,g\rangle,|e,e\rangle\}$, the resulting density matrix is

$$\hat{\rho}_{A,R} = \frac{1}{2}\begin{pmatrix} \frac{1}{Y} & 0 & 0 & \frac{1}{X} \\ 0 & \frac{|\eta_0|^2}{Y} & \frac{-\eta_0^*\eta_1}{X} & 0 \\ 0 & \frac{-\eta_1^*\eta_0}{X} & \frac{|\eta_1|^2}{Z} & 0 \\ \frac{1}{X} & 0 & 0 & \frac{1}{Z} \end{pmatrix}, \tag{12}$$

where

$$X = \sqrt{\left(1+|\eta_0|^2\right)\left(1+|\eta_1|^2\right)},$$
$$Y = 1+|\eta_0|^2, \tag{13}$$
$$Z = 1+|\eta_1|^2.$$

This form of $\hat{\rho}_{A,R}$ is exact within the first-order perturbative treatment of the detector– field coupling: because each branch in Eq. (10) was normalized individually, no further truncation in $\lambda$ is required when constructing Eq. (12), and unitarity is preserved exactly at the level of the truncated state. The diagonal entries $1/(2Y)$ and $1/(2Z)$ give the probability of finding the system in $|g,g\rangle$ and $|e,e\rangle$, respectively, while $|\eta_0|^2/(2Y)$ and $|\eta_1|^2/(2Z)$ give the probability of finding it in $|g,e\rangle$ and $|e,g\rangle$; the off-diagonal entries $1/(2X)$ quantify the coherence surviving between the $|g,g\rangle$ and $|e,e\rangle$ branches, while $-\eta_0^*\eta_1/(2X)$ quantifies the coherence generated between $|g,e\rangle$ and $|e,g\rangle$ by the detector–field interaction. In the limit $\eta_0,\eta_1 \to 0$ (no acceleration, or vanishing coupling), $X,Y,Z \to 1$ and $\hat{\rho}_{A,R}$ reduces to the density matrix of the original Bell state $|\psi_0\rangle$, as required.

### 2.6 Entanglement measure: concurrence

To quantify the entanglement between Alice and Rob, we employ the Wootters concurrence [22], defined as

$$C(\hat{\rho}_{AR}) = max\left\{0, \sqrt{\mu_1}-\sqrt{\mu_2}-\sqrt{\mu_3}-\sqrt{\mu_4}\right\}, \tag{14}$$

where $\sqrt{\mu_1} \geq \sqrt{\mu_2} \geq \sqrt{\mu_3} \geq \sqrt{\mu_4}$ are the eigenvalues of the matrix $\tilde{\rho} = \hat{\rho}_{AR}\left(\hat{\sigma}_y \otimes \hat{\sigma}_y\right)\hat{\rho}^*_{AR}\left(\hat{\sigma}_y \otimes \hat{\sigma}_y\right)$. Using Eqs. (12) and (13) together with Eq. (14), the entanglement can be evaluated numerically as a function of the acceleration $a$ and the detector energy gap $\Omega$.

Fig. 1 shows the concurrence as a function of $a$ for two representative values of $\Omega$. In Fig. 1(a), corresponding to small $\Omega$, the entanglement increases monotonically with acceleration throughout, consistent with the anti-Unruh regime: the effective temperature perceived by the detector decreases with $a$, reducing the associated thermal contribution to detector noise and thereby favoring the preservation of Alice–Rob coherence. In Fig. 1(b), corresponding to larger $\Omega$, the behavior is non-monotonic: At low accelerations, the concurrence decreases due to the detector–field interaction, while at higher accelerations, the anti-Unruh regime reduces the effective temperature, allowing the concurrence to recover and increase. It should be noted that at accelerations close to zero, the concurrence already starts from a value slightly below unity. This reduction does not originate from the Unruh effect but from the detector–field coupling itself: the Gaussian switching activates the interaction even at $a = 0$, introducing a small nonzero $\eta_0$, $\eta_1$ proportional to the coupling strength $\lambda$. The magnitude of this initial reduction therefore scales with $\lambda$.

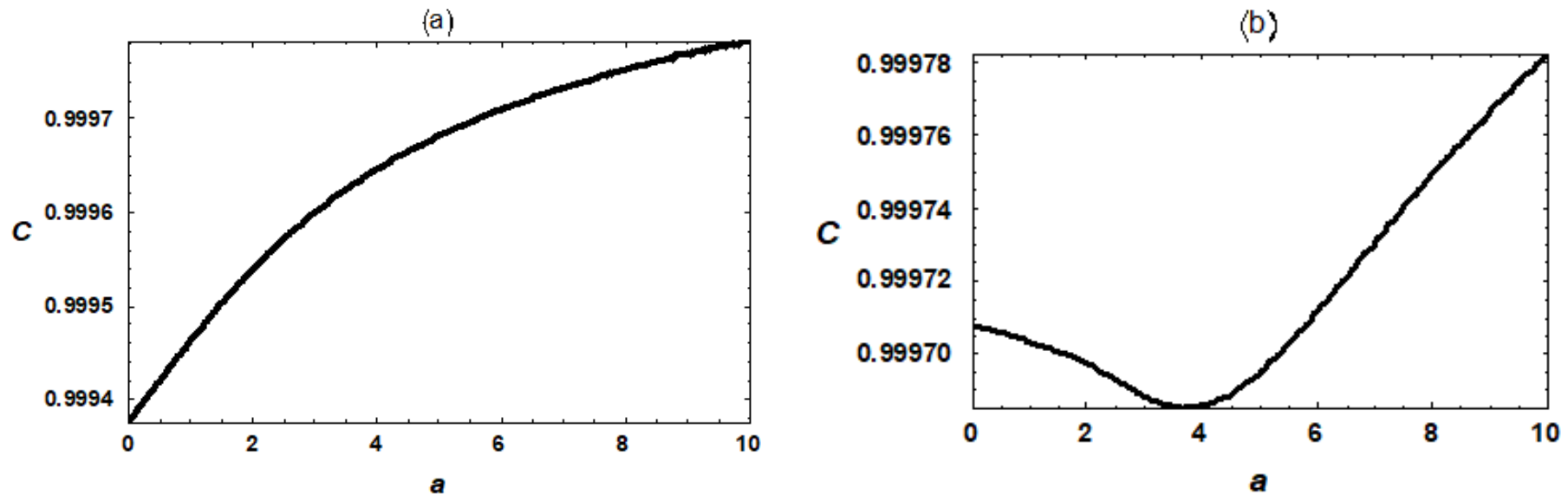

**Fig. 1** Wooster's concurrence C between Alice and Rob as a function of Rob's proper acceleration $a$, for (a) $\Omega = 0.6$ and (b) $\Omega = 6$, assuming $\lambda = 0.06$ and $\sigma = 0.4$.

## 3 Teleportation via accelerated Unruh–DeWitt detectors

### 3.1 Protocol setup

We consider the teleportation of an unknown qubit using two UDW detectors as the physical qubits, with the identification $|g\rangle \equiv |0\rangle$ and $|e\rangle \equiv |1\rangle$. Alice wishes to send the unknown state

$$|\psi\rangle_A = a|0\rangle_A + b|1\rangle_A, \quad |a|^2 + |b|^2 = 1, \tag{15}$$

to Rob. The entangled resource shared between Alice and Rob is prepared while both parties are inertial, in the maximally entangled Bell state

$$|\beta_{00}\rangle_{A,R} = \frac{1}{\sqrt{2}}\left(|0\rangle_A|0\rangle_R + |1\rangle_A|1\rangle_R\right). \tag{16}$$

We emphasize that this resource is prepared before Rob accelerates; preparing it after Rob has already begun accelerating would correspond to a physically distinct initial condition and must be treated separately. With the scalar field initially in the Minkowski vacuum $\left|0\right\rangle_{field}$ the global initial state at preparation time is

$$\left|G\right\rangle = \left|\psi\right\rangle_A \otimes \left|\beta_{00}\right\rangle_{A,R} \otimes \left|0\right\rangle_{field}. \tag{17}$$

Only after $\left|G\right\rangle$ has been prepared in the inertial frame does Rob begin to accelerate Uniformly.

3.2 Bell measurement and classical communication

Alice first applies a CNOT gate to her two qubits, with the input qubit as control and her half of the Bell pair as target, bringing the state to

$$\begin{aligned}\left|G\right\rangle = \frac{1}{\sqrt{2}}\Big(&a\left|0\right\rangle_A\left|0\right\rangle_A\left|0\right\rangle_R + a\left|0\right\rangle_A\left|1\right\rangle_A\left|1\right\rangle_R + b\left|1\right\rangle_A\left|1\right\rangle_A\left|0\right\rangle_R \\ &+ b\left|1\right\rangle_A\left|0\right\rangle_A\left|1\right\rangle_R\Big)\otimes\left|0\right\rangle_{field}.\end{aligned} \tag{18}$$

She then applies a Hadamard gate to the input qubit, yielding

$$\begin{aligned}|G\rangle = \frac{1}{2}\Big(&|00\rangle_A\left(a|0\rangle_R + b|1\rangle_R\right) + |01\rangle_A\left(a|1\rangle_R + b|0\rangle_R\right) \\ &+ |10\rangle_A\left(a|0\rangle_R - b|1\rangle_R\right) + |11\rangle_A\left(a|1\rangle_R - b|0\rangle_R\right)\Big)\otimes|0\rangle_{field}\end{aligned} \tag{19}$$

Alice then measures her two qubits in the computational (Bell) basis. Conditioned on her outcome $(i,j)\in\{00,01,10,11\}$, Rob's qubit collapses to

$$\left|\Phi_{i,j}\right\rangle = \left(x_{i,j}\left|0\right\rangle_R + y_{i,j}\left|1\right\rangle_R\right)\otimes\left|0\right\rangle_{field}, \tag{20}$$

with the four possibilities listed in Table 1.

**Table 1** Rob's qubit state $(x_{i,j}, y_{i,j})$ conditioned on Alice's measurement outcome $(i,j)$.

| Alice outcome $(i,j)$ | Rob's state $(x_{i,j}, y_{i,j})$ |
|---|---|
| 00 | $(a,b)$ |
| 01 | $(b,a)$ |
| 10 | $(a,-b)$ |
| 11 | $(-b,a)$ |

### 3.3 Corrective unitary and received state

After receiving Alice's two classical bits $(i,j)$, Rob applies the corresponding corrective unitary $Z^i X^j$ to recover the original state $|\psi\rangle$. In the standard teleportation protocol [23], these are the usual Pauli gates. However, because Rob is in a non-inertial frame, his qubit is entangled with the field as a result of the Unruh effect; consequently, the action of the corrective unitary must be implemented in conjunction with the transformations of Eq. (6). The state Rob obtains therefore takes the form

$$\begin{aligned}|\Phi_{i,j}\rangle &= \frac{x_{i,j}}{\sqrt{1+|\eta_0|^2}}\left(|g\rangle|0\rangle_{field} - i\eta_0\,|e\rangle\,|1_k\rangle_{field}\right)\\ &+ \frac{y_{i,j}}{\sqrt{1+|\eta_1|^2}}\left(|e\rangle|0\rangle_{field} + i\eta_1\,|g\rangle\,|1_k\rangle_{field}\right),\end{aligned} \tag{21}$$

where $\eta_0$ and $\eta_1$ are the excitation and de-excitation amplitudes introduced in Eq. (7), evaluated for the same acceleration $a$, switching width $\sigma$, and field mass $m$ as in Sec. 2. Tracing Eq. (21) over the field modes, the state received by Rob is the mixed state

$$\rho_{i,j}^{Rob} = \begin{pmatrix} A & B \\ B^* & C \end{pmatrix}, \tag{22}$$

with

$$\begin{aligned} A &= \frac{|x_{i,j}|^2}{1+|\eta_0|^2} + \frac{|y_{i,j}|^2|\eta_1|^2}{1+|\eta_1|^2},\\ B &= \frac{y_{i,j}^* x_{i,j} - x_{i,j}^* y_{i,j}\eta_1\eta_0^*}{\sqrt{\left(1+|\eta_0|^2\right)\left(1+|\eta_1|^2\right)}},\\ C &= \frac{|x_{i,j}|^2|\eta_0|^2}{1+|\eta_0|^2} + \frac{|y_{i,j}|^2}{1+|\eta_1|^2}. \end{aligned} \tag{23}$$

### 3.4 Average teleportation fidelity

To assess the quality of teleportation we use the fidelity [24]

$$F = \langle\psi_{in}|\rho_{out}|\psi_{in}\rangle, \tag{24}$$

where $|\psi_{in}\rangle$ is the pure input state and $\rho_{out}$ is the mixed output state delivered by the channel. To average over all possible $(x_{i,j}, y_{i,j})$, we parametrize

$$x_{i,j} = \cos\left(\frac{\theta}{2}\right), \quad y_{i,j} = e^{i\varphi}\sin\left(\frac{\theta}{2}\right), \tag{25}$$

with $0 \le \theta \le \pi$ and $0 \le \varphi \le 2\pi$, and average using [25]

$$F_A = \frac{1}{4\pi}\int_0^{2\pi} d\varphi \int_0^{\pi} F \sin(\theta) d\theta. \tag{26}$$

Carrying out this average using Eqs. (22) to (26) yields the closed-form result

$$F_A = \frac{3|\eta_1|^2 + 2X + |\eta_0|^2\left(3 + 2|\eta_1|^2\right) + 4}{6X^2}, \tag{27}$$

where $X = \sqrt{\left(1+|\eta_0|^2\right)\left(1+|\eta_1|^2\right)}$ is the same quantity defined in Eq. (13). It should be emphasized that the switching width enters the teleportation fidelity through the transition amplitudes $\eta_0$ and $\eta_1$. Consequently, changing the interaction timescale can modify both the magnitude of the detector induced noise and the acceleration at which the anti-Unruh behavior becomes relevant. In particular, finite time switching may lead to a nonzero detector response even at $a = 0$. In the present work, we fix $\sigma = 0.4$ in order to isolate the acceleration dependence and focus on the operational formulation of the teleportation protocol. A systematic study of $F_A\left(a,\sigma\right)$ is left for future work.

Two limits of Eq. (27) are instructive. In the limit $|\eta_0|,|\eta_1| \to 0$ (vanishing coupling or vanishing acceleration), $X \to 1$ and $F_A \to 1$: the channel becomes noiseless, as it must, since the detector–field interaction is then switched off entirely. Conversely, for strong coupling such that $|\eta_0|,|\eta_1| \gg 1$, $X$ grows as $|\eta_0||\eta_1|$ and $F_A \to 1/3$; this regime, however, lies far outside the perturbative domain $\lambda \ll 1$ in which Eq. (27) was derived, and is never approached for the parameter values considered here. Throughout the parameter range explored in Fig. 2, $F_A$ remains far above the classical teleportation bound $F_{cl} \to 2/3$ [26], confirming that the quantum channel retains a strong advantage over any measure-and-prepare strategy even when degraded by the Unruh effect.

A more quantitative connection to Sec. 2 follows from expanding Eq. (27) to leading order in $|\eta_0|^2$ and $|\eta_1|^2$:

$$1 - F_A \simeq \frac{|\eta_0|^2 + |\eta_1|^2}{3} + O(\eta^4). \tag{28}$$

The infidelity is thus, to leading order, directly proportional to the total excitation probability of Rob's detector, the same combination that controls the off-diagonal suppression of the Alice–Rob density matrix in Eq. (12). This establishes a direct quantitative link between the entanglement degradation discussed in Sec. 2 and the fidelity loss reported here: any mechanism that reduces $|\eta_0|^2 + |\eta_1|^2$, in particular the anti-Unruh effect at large acceleration, reduces the infidelity by the same proportion.

### 3.5 Time-independence of the fidelity

It is known that the free time evolution of Alice's and Rob's states has no effect on the teleportation fidelity in the presence of the Unruh effect alone [2]. We now verify that this property persists in the anti-Unruh regime. The time dependence enters through the free evolution of the detector's ladder operators,

$$\sigma^{\pm}(\tau) = e^{iH_0\tau}\sigma^{\pm}e^{-iH_0\tau} = e^{\pm i\Omega\tau}\sigma^{\pm}, \tag{29}$$

where $H_0 = \Omega\sigma_z/2$ is the detector's free Hamiltonian. Under this time evolution, the state of Eq. (20) evolves as

$$\left|\Phi_{i,j}\right\rangle = \left(e^{i\frac{\Omega}{2}\tau}x_{i,j}\left|0\right\rangle_R + e^{-i\frac{\Omega}{2}\tau}y_{i,j}\left|1\right\rangle_R\right)\otimes\left|0\right\rangle_{field}. \tag{30}$$

After discarding the overall global phase, which has no physical effect, this can equivalently be written as

$$\left|\Phi_{i,j}\right\rangle = \left(x_{i,j}\left|0\right\rangle_R + e^{-i\Omega\tau}y_{i,j}\left|1\right\rangle_R\right)\otimes\left|0\right\rangle_{field}. \tag{31}$$

Repeating the partial trace over the field modes and the averaging procedure of Eq. (26), the phase $e^{-i\Omega\tau}$ drops out identically, and Eq. (27) is recovered unchanged. The fidelity is therefore independent of the local free-evolution time of the detectors, both in the Unruh and the anti-Unruh regimes.

### 3.6 Results: fidelity versus acceleration

Fig. 2 illustrates the average fidelity as a function of the acceleration $a$ for the same parameter sets as in Fig. 1. In panel (a), corresponding to small $\Omega$, $F_A$ increases monotonically with $a$ throughout, mirroring the monotonic growth of the concurrence in Fig. 1(a): in this regime the anti-Unruh effect dominates at all accelerations, so $|\eta_0|^2 + |\eta_1|^2$ decreases monotonically and, by Eq. (28), $F_A$ rises correspondingly.

In panel (b), corresponding to larger $\Omega$, the fidelity exhibits a much weaker signature of the underlying Unruh–anti-Unruh crossover than the concurrence does in Fig. 1(b): $F_A$ remains nearly flat (with only a shallow minimum) at low-to-intermediate accelerations, before rising more steeply once the anti-Unruh effect becomes dominant. This contrast between Figs. 1(b) and 2(b) follows directly from the different functional dependence of $C$ and $F_A$ on $\eta_0, \eta_1$: the concurrence is built from eigenvalues of the full 4 × 4 density matrix and is sensitive to the detailed coherence structure between $\left|g,e\right\rangle$ and $\left|e,g\right\rangle$, whereas $F_A$, being an average over all input states [Eq. (26)], partially washes out this sensitivity, so that the Unruh-induced dip is comparatively less pronounced. In both panels, the high-acceleration limit $F_A \to 1$ reflects the same physical origin established in Sec. 2: the anti-Unruh suppression of $|\eta_0|^2$ and $|\eta_1|^2$ drives the teleportation channel toward the noiseless limit.

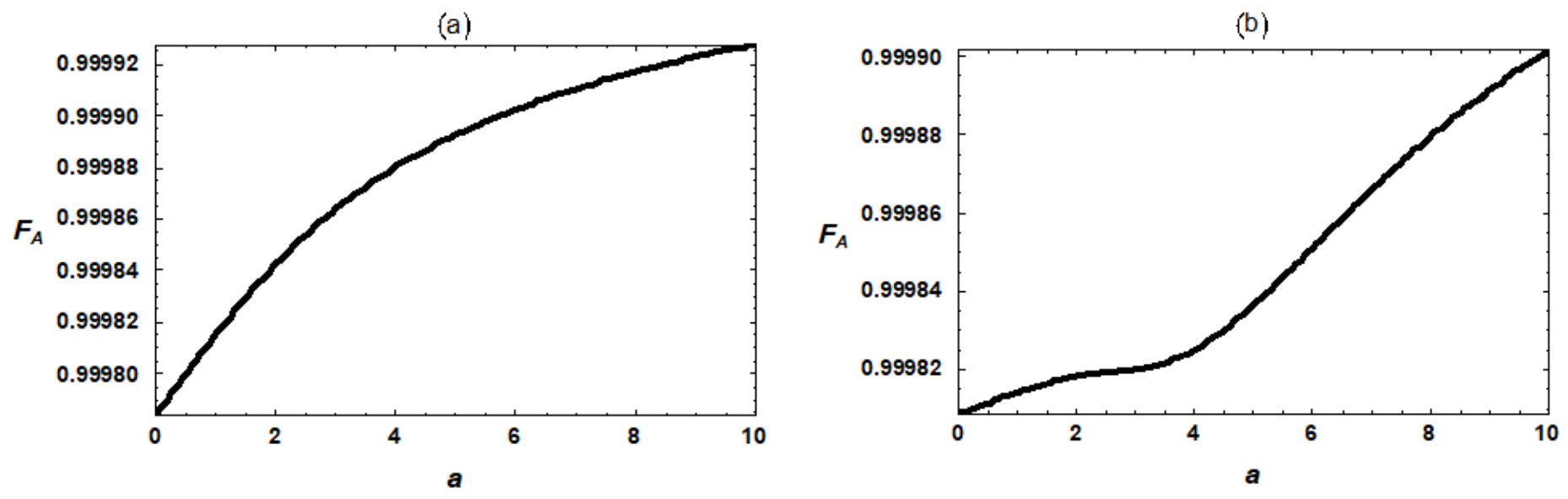

**Fig. 2** Average teleportation fidelity as a function of Rob's proper acceleration $a$, for (a): $\Omega = 0.6$ and (b): $\Omega = 6$. Parameters are the same as in Fig.1.

### 3.7 Quantum information recovery: entropy analysis

We next examine Rob's von Neumann entropy at two stages of the protocol: before Alice's measurement result is communicated, and after Rob has received the classical bits $(i, j)$ and applied the corrective unitary $Z^i X^j$. This comparison is of particular interest because, in the presence of the Unruh effect alone, the difference between these two entropies is known to decrease with increasing acceleration [2]; we expect the opposite trend, an increase in this difference, once the anti-Unruh effect becomes dominant. The von Neumann entropy is defined as [27]:

$$S(\rho) = -\sum_i \lambda_i \log_2(\lambda_i), \tag{32}$$

where $\lambda_i$ are the eigenvalues of $\rho$.

It should be noted that, before Rob undergoes acceleration and his detector interacts with the massive field, his reduced density matrix is $\rho_R^{pre} = \frac{1}{2} I_{2\times 2}$; the corresponding von Neumann entropy is therefore equal to unity, indicating that Rob initially has no information about the quantum state that Alice intends to teleport. After the detector–field interaction and the onset of acceleration, Rob's reduced density matrix is obtained by tracing over Alice's subsystem in the total density matrix of Eq. (12). Following the reception of Alice's classical information, Rob's reduced density matrix is given by Eq. (22). The information gained by Rob is quantified by

$$\Delta S = S_R^{pre} - S_R^{\mathrm{Reception}}. \tag{33}$$

Table 2 lists $S_R^{\mathrm{Reception}}$ and $\Delta S$ for several values of the acceleration; the intermediate entropy $S_R^{pre}$ is omitted from the table because, to the numerical precision reported here, it remains equal to unity for all accelerations considered. Before reception, Rob's entropy is thus always close to its maximal value, reflecting near-complete uncertainty about Alice's state prior to any classical communication. After reception, $S_R^{\mathrm{Reception}}$ decreases with acceleration, eventually approaching zero at very high accelerations, indicating that Rob obtains a complete quantum bit. The entropy difference $\Delta S$, listed in the final column, represents the amount of information (in bits) successfully transferred to Rob: at low accelerations Rob receives slightly less than one quantum bit, while at very high accelerations he receives a full quantum bit. This demonstrates that information initially lost to the field via the detector–field interaction is recovered at high accelerations as a consequence of the anti-Unruh effect.

**Table 2** Rob's entropy after receiving the information $\left(S_R^{\mathrm{Reception}}\right)$ and the amount of information received $\left(\Delta S = S_R^{pre} - S_R^{\mathrm{Reception}}\right)$, for several values of the acceleration.

| $a$ | $S_R^{\mathrm{Reception}}$ | $\Delta S$ |
|---|---|---|
| $10^{-3}$ | 0.00411 | 0.99589 |
| 1 | 0.00358 | 0.99642 |
| $10^3$ | 0.00001 | 0.99999 |
| $10^6$ | $\approx 0$ | $\approx 1$ |

## 4 Conclusion

In this work, we have reformulated the standard quantum teleportation protocol in a relativistic scenario where one of the participants, Rob, undergoes uniform acceleration while Alice remains inertial. Rob's subsystem was modeled as an Unruh–DeWitt detector locally coupled to a massive scalar field. Within this framework, we analytically derived the resulting quantum channel and the quantum information transferred to Rob. We then analyzed the behavior of entanglement, teleportation fidelity, and entropy to clarify how the detector–field interaction and its acceleration dependence affect the flow of quantum information.

Our results show that, for small detector energy gaps, the shared entanglement between Alice and Rob increases monotonically with acceleration. For larger energy gaps, entanglement initially decreases at low accelerations due to the acceleration-dependent detector–field dynamics, but is subsequently restored and enhanced at higher accelerations as the anti-Unruh effect emerges. This transition is accompanied by a reduction in the detector's effective temperature, which can reduce the associated decoherence and enable a partial recovery of quantum coherence previously lost to the field.

These features are directly reflected in the performance of the teleportation protocol. Although the interaction with the quantum field leads to an initial loss of information and an increase in Rob's entropy, the onset of the anti-Unruh effect at higher accelerations reverses this trend. As a consequence, the teleportation fidelity increases with acceleration and approaches unity in the high-acceleration regime, indicating an effective recovery of the quantum information initially dispersed into the field. We also find that, the teleportation fidelity is independent of the local free-evolution time of Alice's and Rob's states.

Our findings demonstrate that the anti-Unruh effect plays a constructive role in relativistic quantum information processing, in contrast to the commonly emphasized destructive impact of the Unruh effect. In particular, the recovery of entanglement and information observed here shows that the anti-Unruh regime can help preserve quantum correlations and quantum information in a non-inertial frame. This work highlights the relevance of anti-Unruh phenomena for relativistic quantum communication protocols and provides further insight into the interplay between detector–field interactions, quantum correlations, and information flow in non-inertial frames.

**Acknowledgements** The authors thank Shahid Chamran University of Ahvaz for their support.

**Funding** This research was funded by Shahid Chamran University of Ahvaz (Grant Numbers SCU.SP1404.812 and SCU.SP1404.479).

**Data Availability Statement** No datasets were generated or collected during this study. All results are derived from analytical formulations presented in the manuscript. Therefore, data sharing is not applicable.

**Code Availability Statement** The custom Mathematica codes developed for numerical computations and figure generation are available from the corresponding author upon reasonable request.

## Declarations

Competing interests The authors declare no conflict of interest.